# A CLIMATE CHANGE VULNERABILITY ASSESSMENT FRAMEWORK: A SPATIAL APPROACH

Claudia Cáceres, Claremont Graduate University, Claremont, USA, claudia.caceres@cgu.edu

Yan Li, Claremont Graduate University, Claremont, USA

Brian Hilton, Claremont Graduate University, Claremont, USA

**Abstract:** Climate change is affecting every known society, especially for small farmers in Low-Income Countries because they depend heavily on rain, seasonality patterns, and known temperature ranges. To build climate change resilient communities among rural farmers, the first step is to understand the impact of climate change on the population. This paper proposes a Climate Change Vulnerability Assessment Framework (CCVAF) to assess climate change vulnerabilities among rural farmers. The CCVAF framework uses information and communication technology (ICT) to assess climate change vulnerabilities among rural farmers by integrating both community level and individual household level indicators. The CCVAF was instantiated into a GIS-based web application named THRIVE for different decision-makers to better assess how climate change is affecting rural farmers in Western Honduras. Qualitative evaluation of the THRIVE showed that it is an innovative and useful tool. The CCVAF contributes to not only the knowledge base of the climate change vulnerability assessment but also the design science literature by providing guidelines to design a class of climate change vulnerability assessment solutions.

## 1. INTRODUCTION

According to the Intergovernmental Panel on Climate Change (IPCC) Fifth Assessment Report (AR5), climate change has a clear human influence with the highest anthropogenic greenhouse gas emissions (GHG) in history, diminishing snow levels and ice caps, rising sea levels, and warming atmosphere and oceans. Disadvantaged people, such as rural poor and smallholder producers in developing countries, are at a higher risk of climate change as the changes in climate patterns will impact crop yields and undermine food security, especially among subsistence farmers who generally produce low yields and are least able to cope with their effects (Altieri et al., 2015; Antle, 1995; FAO, 2017; IPCC, 2014; P. Jones & Thornton, 2003; Kang et al., 2009; Misra, 2014; Schmidhuber & Tubiello, 2007; UN, 2018; World Bank, 2013). Thus, they are a priority in climate change adaption plans (Holland et al., 2017; Morton, 2007).

Climate change adaptation focuses on strengthening resilience and reducing vulnerability (FAO, 2018). It should involve the local communities, civil society, international organizations, governments at the local, regional and national levels (Neil Adger et al., 2005). They need to assess these populations' adaptive capacity and identify vulnerabilities (Bouroncle et al., 2017). Information is often limited due to the difficulty of obtaining data about these vulnerable populations, and their expected shocks and stresses, particularly those faced by marginal communities of small farmers in low-income countries (LICs). To help build climate change resilient communities among rural farmers, the first step is to understand the impact of climate change on the population, its land, and its agricultural practices. This research aims to use information, communication, and technology (ICT) to assess climate change vulnerabilities among rural farmers. To achieve this overall objective, this research seeks to answer two research questions:

1. What determinants and indicators are needed to assess climate change vulnerability?
2. How can these indicators be measured and integrated to assess climate change vulnerabilities among rural farmers?

The rest of the paper is organized as follows. Section 2 presents the literature review, followed by the research methodology in Section 3. Section 4 describes the CCVAF, and Section 5 presents a Geographic Information Systems (GIS)-based web application that instantiates the framework using a case study in Western Honduras. Section 6 evaluates the CCVAF framework, and its instantiation, and Section 7 concludes the paper.

## 2. LITERATURE REVIEW

The concept of Resilience has seen an evolution within the climate change and development community as the "capacity for adaptation, learning, and transformation" (L. Jones et al., 2019a). Resiliency is how a system can absorb risk and react to hazard through time. Adaptive capacity is the adaptability of a system to reduce its vulnerability (IPCC & Edenhofer, 2014; McCarthy & IPCC, 2001). Adaptive capacity needs the evaluation and measure of different characteristics inherent in its system to identify possible solution paths and tools to mitigate and cope with risk. The adaptive capacity has been viewed as a part of a wider resilient model (L. Jones et al., 2019a; Sorre et al., 2017a). The three ways to build resiliency in communities are: a) reduce exposure, b) reduce sensitivity, and c) increase adaptive capacity (Meybeck et al., 2012).

Vulnerability describes the analysis to measure powerlessness, marginality, and how susceptible a group or individual can be to a harmful situation being caused by multiple stressors and pathways (Adger, 2006). It has become a central concept to climate change research as its effects are being widely observed and the development of vulnerability assessments are being used to raise awareness, develop policies, and monitor adaptation measures (GIZ, 2013, 2014; Hinkel, 2011). If one intends to create a vulnerability assessment (to encourage a change in a community or inform policymakers), one must determine the methodology to measure vulnerability. Empirical studies show the use of a variation of the basic formula to measure vulnerability: "Vulnerability = Risk + Response" or "Vulnerability= Baseline + Hazard + Response" (Moret, 2014).

Different approaches can be used to build vulnerability assessments as the three ontological approaches identified by (Below et al., 2012a): theory-driven, data-driven, and combination of empirical and theoretical. The theory-driven approach uses a literature review to select the variables being measured, but this approach provides a level of uncertainty as to whether the chosen variables can measure vulnerability. The data-driven approach selects the variables being measured through expert opinion or the correlation of past events, but this approach does not assess the variables through a benchmark but limits itself to expert opinion. The third approach is a response to the weaknesses of the other approaches. Two specific examples are the Livelihood Vulnerability Index proposed by (Hahn et al., 2009) and the Vulnerability assessment using an Indicator approach proposed by (Below et al., 2012b; Gbetibouo et al., 2010). The indicator approach uses specific indicators or a combination of them to measure vulnerability to compute indices or weighted averages (Gbetibouo et al., 2010). To the best of our knowledge, there is no vulnerability assessment approach towards climate change.

## 3. METHODOLOGY

To answer the research questions, this research targets the designing and development of a framework to integrate vulnerability indicators and their measure to assess climate change vulnerabilities among rural farmers. Thus, this research undertakes a design science research paradigm (Hevner et al., 2004; Hevner & Chatterjee, 2010), as it designs and implements innovative and useful artifacts to solve real-world problems.

The first artifact is a Climate Change Vulnerability Assessment Framework (CCVAF) using ICT, specifically GIS-based remote sensing. The design of CCVAF includes the understanding of the problem and then acquiring knowledge from the environment within its design to solve it. Thus, it is important to interact with the people and organizations where the CCVAF will be implemented (Hevner et al., 2004). Thus, the researchers maintained a close interaction for several months with the THRIVE (Transforming Household Resilience in Vulnerable Environments) team from World Vision (WV), a global humanitarian organization. Such a close interaction allowed the researchers to better understand the practitioners' needs and processes on the vulnerability assessment based on the data collected. Their feedback was essential in the iterative development of the CCVAF.

As an abstract framework, its validation needs to be tested through its implementation (Hevner & Chatterjee, 2010; Nunamaker Jr. et al., 1990). To instantiate the proposed CCVAF, a web-based application named THRIVE was developed for the THRIVE team, focusing on Western Honduras, the Departments (i.e., regional governments) of Intibucá, Lempira, Ocotepeque, Copan, and Santa Barbara with a total area of 17,303.13 km$^2$ and 114 municipalities. Honduras, a small low-middle-income country with more than 60.9% of its population living in poverty and one out of five Hondurans from rural communities living in extreme poverty (i.e., less than US$2.00 per day)(Ben-Davies et al., 2014; World Bank, 2018).

Artifact evaluation is an important step to demonstrate its utility. THRIVE was evaluated through a qualitative approach to understanding its usefulness and ease of use. The qualitative method used semi-structured interviews as data collection methods and was conducted through Zoom and Teams.

## 4. FRAMEWORK

The Climate Change Vulnerability Assessment Framework (CCVAF) is shown in Figure 1. It includes four steps: 1) identify vulnerability indicators using a hierarchical approach, 2) Identify data sources and collection, including GIS-based remote sensing; 3) measure the vulnerability indicators through GIS-based data analysis and modeling approach, and 4) create an overall index for areas of interests and visualize the results.

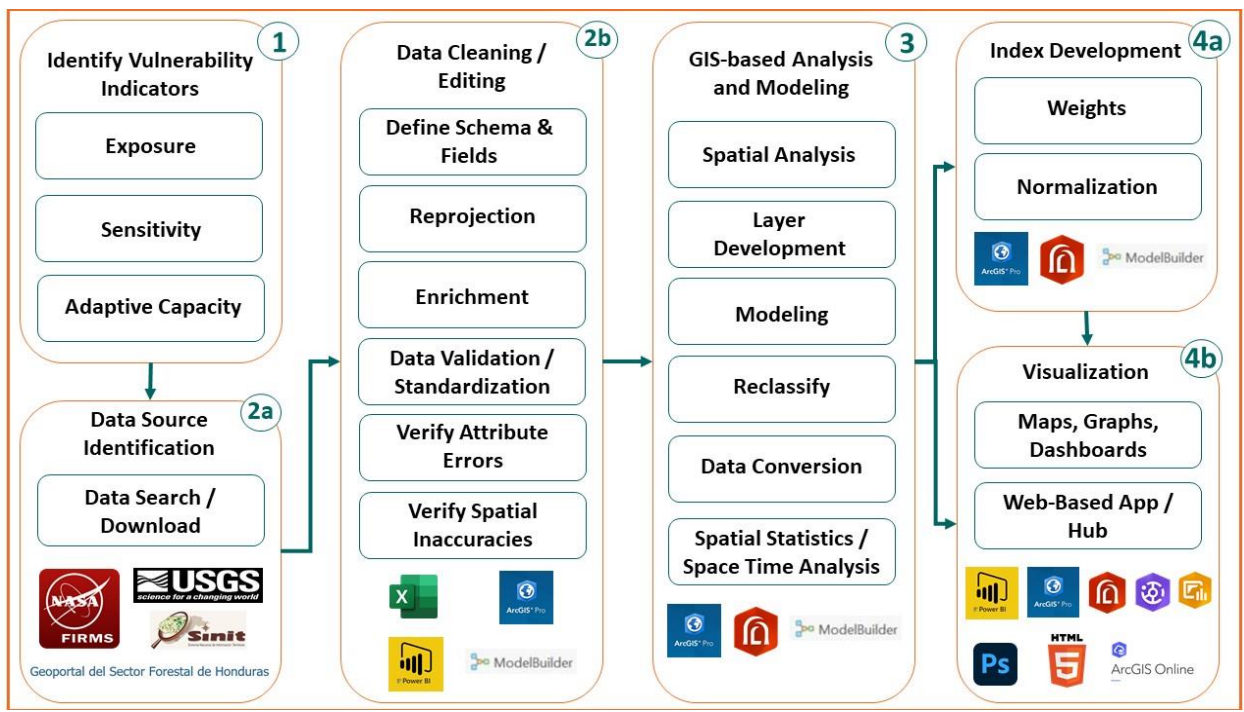

*Figure 1:* **CCVAF and its detailed steps**

The first step is to identify the vulnerability indicators, which answer the first research question. Table 2 lists a comprehensive set of indicators for climate change vulnerability assessment. These indicators were adapted from the research of (Banu et al., 2011; Below et al., 2012b; Caceres, 2011; Gbetibouo et al., 2010; Hahn et al., 2009; Hirschi et al., 2011; Jaiswal et al., 2002; Shah et al., 2013), including the methodology used by UNDP to measure Unsatisfied Basic Needs as developed in CEPAL & PNUD (1988b). Depending on the area of the study, the practitioners and researchers may only select a subset of these indicators that are relevant to their study objectives. For example, in the THRIVE app, indicators related to the economic capacity, financial and market access from the adaptive capacity determinant were not included due to the COVID-19 travel restrictions.

The second step focuses on how to collect and process data in a format for analysis and modeling. As shown in Table 2, many measurements for adaptive capacity are straightforward to process, while these related to the exposure and sensitivity rely heavily on GIS and remote sensing data that requires an additional process. These data have been extensively used to perform complex spatial analysis to mitigate climate change impacts, such as identifying fire risk zones (Jaiswal et al., 2002), measuring environmental degradation (Hassan et al., 2015), estimating crop productivity (Tan & Shibasaki, 2003), and developing climate adaptation model tools (Kunapo et al., 2016). The data pre-processing process may include enrichment, reprojection, and cleaning, and may require different tools such as Microsoft Excel, Power BI, and ArcGIS Pro.

The third step starts with the creation of a database to store geospatial data or geodatabase. It may include data selection, filtering, querying, creation, and export to a different format. For raster data, it may be necessary to create a mosaic for image classification. Raster data need to be converted to vector and vector data may need to be converted to raster for analysis. Tools for data analysis may include ESRI ModelBuilder, a visual programming language inside ArcGIS Pro to build geoprocessing workflows (ESRI, 2020). Short Arcade or Python scripts may also be used to perform calculations.

The last step follows the guidelines presented by ((USAID (2), 2014) as best practices in the composite indices. The creation of the index includes three steps. First, based on the indicators for each subcomponent, an overall index for each subcomponent is calculated. The overall index uses:

$$Index = \Sigma W_1 X_1 + \cdots + W_n X_n$$

where $W_1$ = weight factor and $X_1$ = indicator.

The second step will determine the weight for each component, and then calculate the overall index for each component using normalized subcomponent values. Finally, an overall index to calculate vulnerability using:

$$VI = W_{x1} I_{s1} + \cdots + W_{xn} I_{sn}$$

where W can be defined by the users based on their specific decision context.

We illustrate the development of the Fire Risk Index and the Vulnerability Index using the Western Honduras data. A Fire Risk Index integrates several variables: 1) topographic variables (slope, elevation, and aspect) 2) socioeconomic variables (settlements and roads), and 3) land cover. But through literature review and expert advice, it was determined some variables have a higher influence regarding fire risk. A schematic model was developed using the variables. The Fire Risk Index formula can be summarized as follows:

$$\text{Fire Risk Index} = 1 + 75lc + 30sl + 10a + 5r + 5se + 2e$$

Where *lc* is the Land Cover score, *sl* is the Slope Layer score, r is the proximity to Road score, *se* is the proximity to settlements score, and et is the elevation score.

**Table 1: Vulnerability Assessment Indicators, and related measurements and data source**

| Vulnerability Determinant | Component | Sub-component | Indicator | Unit of Measurement | Data Source | Source |
|---|---|---|---|---|---|---|
| **Exposure** | Extreme Climate Events | Droughts (Water Scarcity) | Frequency of Droughts | Number of Droughts | GIS/Remote Sensing Analysis | (Gbetibouo et al., 2010) |
| | | Flood | Frequency of Flood | Number of Floods | GIS/Remote Sensing Analysis | (Gbetibouo et al., 2010) |
| | Change in Climate | Change in Temperature | Change in Temperature | Degrees Celsius Change | GIS/Remote Sensing Analysis | (Gbetibouo et al., 2010) |
| | | Change in Precipitation | Change in Precipitation | mm Change | GIS/Remote Sensing Analysis | (Gbetibouo et al., 2010) |
| | Forest Fires | Forest Fires | Forest Fire Risk | Area in Kilometers | GIS/Remote Sensing Analysis | (Caceres, 2011; Jaiswal et al., 2002) |
| | Soil Moisture | Soil Moisture | Change in Soil Moisture | Area in Kilometers | GIS/Remote Sensing Analysis | (Hirschi et al., 2011; Kumar et al., 2018) |
| | Soil Carbon | Soil Organic Carbon | Soil Organic Carbon | Area in Kilometers | GIS/Remote Sensing Analysis | (Angelopoulou et al., 2019; Bhunia et al., 2019; Wang et al., 2013) |
| **Sensitivity** | Deforestation | Change in Land Cover | Change in Land Cover | Kilometers of Land Cover | GIS/Remote Sensing Analysis | (Lawrence & Vandecar, 2015) |
| | Land Degradation Index | Percentage of Land Degradation | Percentage of Area with High Land Degradation Index | No Units | GIS/Remote Sensing Analysis | (Gbetibouo et al., 2010) |
| | % Irrigated Land | Percentage of Irrigated Land | Number of Farms with Irrigation Systems | Number of Farms | Does your farm have any type of irrigation system? | (Gbetibouo et al., 2010) |
| | % Small-Scale Farming Operation | | Percentage of Area with Higher Number of Small-Scale Farming Operations | Percentage | What is the area of your farm? | (Gbetibouo et al., 2010) |
| | Crop Diversification Index | | Number of Crop Types | Percentage | What are the crops on this farm? Do you rotate the crops? | (Gbetibouo et al., 2010) |
| **Adaptive Capacity** | Socioeconomic | Economic Capacity | Number of Household Members | Number of Members | How many members live in this household? | (Below et al., 2012a) |
| | | | Number of Households where the Primary Adult is Female | Number of Households with Female Head | Who is the head of the family? Male or female | (Shah et al., 2013) |
| | | | Number of Years the Head of Household Attended less than 3 Years of School | Years | Did you go to school? If yes, what was the last grade you attended? | (CEPAL & PNUD, 1988; Shah et al., 2013) |
| | | | Number of Heads of Household whose age is under 18 and over 45 | Years | What is the age of the head of household? | (CEPAL & PNUD, 1988) |
| | | | Number of Members in the Household who are Employed | Number of Members | How many members of the household are currently employed? What is the type of occupation? | (CEPAL & PNUD, 1988; Islam & Winkel, 2017) |
| | | | Number of Members Working outside the Community | Number of Members | How many members worked outside the community? | (Hahn et al., 2009) |
| | | | Number of Households Receiving Remittances on a Regular Basis | Number of Households | Do you regularly receive remittances? | (Mochizuki et al., 2014; Rajan & Bhagat, 2017) |
| | | Dependency | Population under 14 and over 60 Years of Age | Ratio of Number of Members | How many members are under 14 and over 60? | (Below et al., 2012a; Hahn et al., 2009) |
| | | | Population with Physical or Mental Disability | Ratio of Number of Members | Is there a member of the household with physical or mental illness or disability? If yes, how many? | (Shah et al., 2013) |
| | | | Number of Households with Orphans | Number of Members | Are there any children over 18 from other families living in this house because on or both of their parents died or moved to another country? | (Hahn et al., 2009) |
| | Access to Basic Sanitary Service | Availability | Source of Water | Kilometers | What is the household's source of water? a) well b) river c) public service d) bottled water truck | (Below et al., 2012a; CEPAL & PNUD, 1988) |
| | | | Distance to the Source of Water | Kilometers | How long do you walk to the source of water? A) 0 b) 0.5 km c) 1 km d) 1.5 km e) 2 km f) more than 2 km | (Below et al., 2012a; CEPAL & PNUD, 1988) |
| | | Sewage Disposal System | Type of Sewage Disposal system | Type of Sewage | What is the type of sewage disposal system? A) toilet connected to sewer b) toilet drains in river c) latrine with septic tank d) common pit latrine e) no basic sanitary service or latrine | (CEPAL & PNUD, 1988) |

| | Financial Access | Access to Credit | Number of Households with Access to Credit | Number of Households | Do you have access to credit? When was the last time you received credit? | (Gbetibouo et al., 2010) |
|---|---|---|---|---|---|---|
| | Market Access & Analysis | Distance to Markets | Distance to Nearest Market | Minutes | How far is the nearest market? | (Below et al., 2012a) |
| | | Quality of Road | Quality of Road | Paved or Unpaved | GIS Analysis | (Gbetibouo et al., 2010) |
| | Health | Chronic Illness | Number of Household Members with a Chronic Illness | Number of Members | How many household members suffer from a chronic illness? | (Hahn et al., 2009) |
| | | Access to Health Service | Number of Households with at least a Basic Health Center in a 5 km radius | Number of Households | GIS Analysis | (Hahn et al., 2009) |
| | | Dengue, Zika, Chikungunya exposure | Number of Household with Bed Nets | Number of Households | Do you have bed nets? | (Hahn et al., 2009) |
| | | | Areas with a High Number of Cases | Area $Km^2$ | GIS Analysis | (Hahn et al., 2009) |
| | | | Number of Members who Experienced Dengue or Similar Episode in the Last Month | Number of Members | How many of your household members suffered from Dengue, etc.? | (Hahn et al., 2009) |
| | Knowledge and Information | Access to Knowledge and Information | Number of Households with Access to Information and Knowledge | Number of Households | Do you have access to a reliable system for climate, weather, land or market information? | (L. Jones et al., 2019b; Sorre et al., 2017b) |
| | | | Number of Local Organizations and Community Leaders with Access to Information and Knowledge | Number of Local Organizations and Community Leaders | Do you have access to a reliable system for climate, weather, land, or market information? | |

## 5. THRIVE APP

The THRIVE web-based app is a GIS-based visualization and knowledge platform (see Figure 2) that aims to support decision-makers, such as NGOs, local government, or policymakers in assessing climate change vulnerabilities among rural farming communities. It includes information on how Climate Change is affecting the region, including forest fire risk zones, deforestation, access to health, and vulnerable areas. It allows users to explore, visualize and export information using the different tools provided.

The app included three determinants to calculate the Vulnerability Index: Exposure, Sensitivity, and Adaptive Capacity. The Exposure tab includes an Introduction Story Map, Hotspot Dashboard, Fire Risk Zones Dashboard, and the Soil Moisture Dashboard. The Sensitivity tab includes an Introduction Story Map, the forest loss and gain dashboard, the forest cover change app, and the agriculture dashboard. The Adaptive Capacity tab includes the Introduction Story Map, the Access to Health dashboard, Access to Basic Housing, Access to Basic Sanitary Services dashboard, and the Dependency Dashboard. The app also includes a tab to visualize the Vulnerability of the area by Department, Municipality, and Village. Several dashboards also include web apps with tools allowing users to print, measure, draw and export the layer table. Every section includes an introduction section to help the user understand the methodology used in the analysis and every dashboard includes a How-To section to help the user navigate throughout the app. The web app was developed using ESRI ArcGIS Online especially Web App Builder and Operation Dashboard.

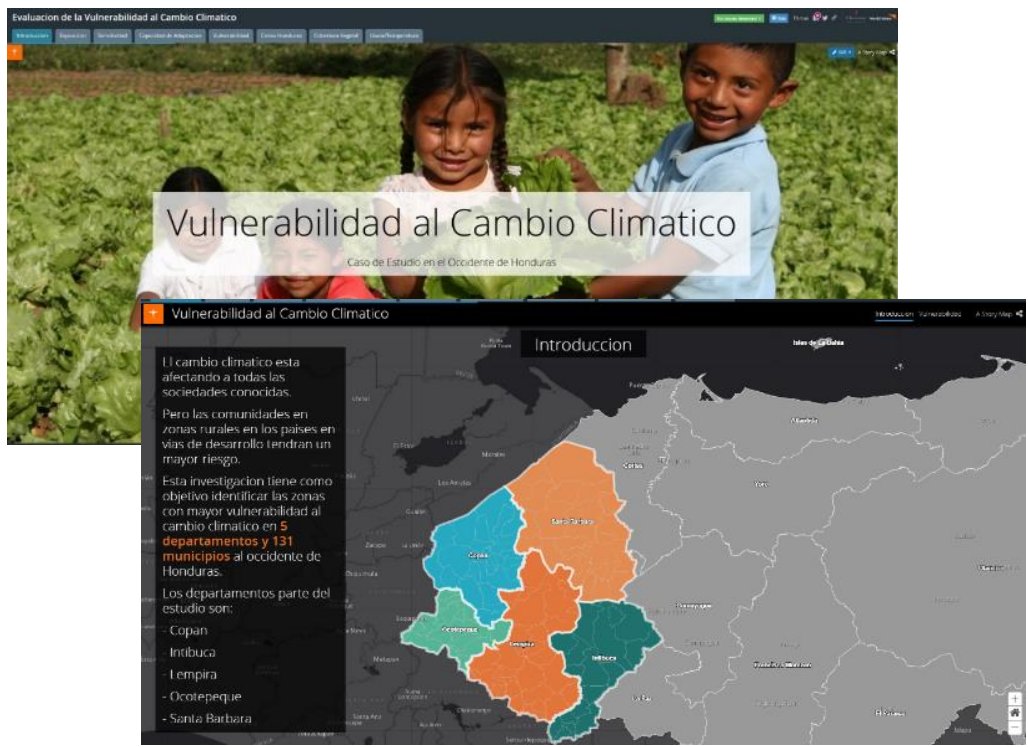

**Figure 2. Initial screens introducing the THRIVE web app.**

## 6. EVALUATION

As described earlier, semi-structured interviews were used to evaluate the THRIVE web app, focusing on its ease of use and usefulness. An initial interview was conducted with the WV Development Officer, followed by six additional interviews with professionals outside World Vision, one located in Honduras and the rest located in the US but able to speak and read Spanish.

All participants considered the web-based app as innovative and useful, as shown below:

Participant 1, professional outside World Vision:

> "If I were a local government or authority in the region, I would see this tool as extremely useful to identify where the population is living, under what conditions, and providing useful insights for decision making."

Participant 2, World Vision staff:

> "This tool will be extremely useful, for example in a project design, very soon we will start the process for the 2021 – 2026 strategy planning, and I believe this tool will play an important role during this process. I see us using it for the climate change transverse axis in our projects, specifically with climate change adaptation processes."

In general, most of the participants considered the tool was easy to use with some exceptions from professionals who mentioned they were initially not sure where to start or what to do.

Participant 1, Professional outside World Vision mentioned:

> "Initially I was very confused, and I didn't know what to do. But once I got into the guide, it was very easy to use."

Participant 2, World Vision staff mentioned:

> "Something I like about this tool, is that it is very easy to use. And if needed the help sections in the left panels provide additional support on how to use it."

Participant 8, World Vision staff mentioned:

> "I consider this tool very user-friendly."

During the evaluation, one of the most common themes was the recommendations from the participants to edit or change sections from the app. All the recommendations provided were used to improve the app. For example, Participant 1, stated:

> "I think it is a little bit disorganized, and if you could organize it better maybe using the determinants used for the analysis. I think the Census tab seems a little bit outside the topic and I think should not be the first tab maybe change its order."

## 7. CONCLUSIONS

This study proposes a Climate Change Vulnerability Assessment Framework (CCVAF) to better evaluate the different indicators for vulnerability assessment. The framework not only allows the assessment of the overall climate change vulnerability but also the understanding of how different vulnerability indicators would impact the overall vulnerability to support decision making in building climate change resilient in communities. A GIS-based web application, named THRIVE, was implemented to instantiate CCVAF. Although the THRIVE app is built specifically for Western Honduras, its design is based on the CCVAF framework and can be easily extended to different areas around the world.

Further research is needed to examine the exposure and sensitivity determinants along with adaptive capacity. For the exposure, several components could be analyzed using extreme climate events, change in climate and soil carbon. For the sensitivity, future research could include the percentage of irrigated land, crop diversification, and land degradation. For adaptive capacity, future research could include measurements of economic capacity and access to basic sanitary service at a household level, financial access, market access, and improved health access. Additionally, a research plan would be developed to include the expansion of THRIVE app to other areas of Honduras and in the Central American region.